\documentclass{imsart}

\usepackage{amsfonts}
\usepackage{amsmath}
\usepackage{amsthm}
\usepackage{dsfont}

\RequirePackage[OT1]{fontenc}
\RequirePackage{amsthm,amsmath,natbib,verbatim}
\RequirePackage[colorlinks=true,citecolor=blue,linkcolor=blue,urlcolor=blue]{hyperref}

\startlocaldefs
\numberwithin{equation}{section}
\theoremstyle{plain}
\newtheorem{thm}{Theorem}[section]
\newtheorem{lem}[thm]{Lemma}

\newcommand{\ind}{{\mathds{1}}}
\DeclareMathOperator{\Prob}{Pr}
\newcommand{\bs}{\boldsymbol}
\endlocaldefs

\begin{document}

\begin{frontmatter}
\title{A Dynamic Programming Approach for Approximate Uniform Generation of Binary Matrices with Specified Margins}
\runtitle{Approximate Uniform Generation of Binary Matrices}

\begin{aug}
\author{Matthew T. Harrison}\thanksref{t1}

\address{Division of Applied Mathematics \\ Brown University \\ Providence, RI 02912 \\
email: {\tt Firstname\_Lastname@Brown.edu}}

\thankstext{t1}{This work was supported in part by the National Science Foundation under Grant DMS-0240019 and by the National Institutes of Health under Grant NIMH-2RO1MH064537 while the author was in the Department of Statistics at Carnegie Mellon University.}
\runauthor{M.T. Harrison}

\affiliation{Carnegie Mellon University}

\end{aug}

\begin{abstract} {} Consider the collection of all binary matrices having a specific sequence of row and column sums and consider sampling binary matrices uniformly from this collection.  Practical algorithms for exact uniform sampling are not known, but there are practical algorithms for approximate uniform sampling.  Here it is shown how dynamic programming and recent asymptotic enumeration results can be used to simplify and improve a certain class of approximate uniform samplers.  The dynamic programming perspective suggests interesting generalizations.
\end{abstract}



\tableofcontents
\end{frontmatter}

\section{Problem summary} \label{s:problem}

Consider the set of $m\times n$ binary matrices with row sums $\bs{r}:=(r_1,\dotsc,r_m)$ and column sums $\bs{c}:=(c_1,\dotsc,c_n)$, namely,
\[ \Omega(\bs{r},\bs{c}) := \textstyle \bigl\{\bs{z}\in\{0,1\}^{m\times n} : \sum_{\ell=1}^n z_{i\ell}=r_i , \ \sum_{\ell=1}^m z_{\ell j}=c_j, \forall i,j \bigr\} \]
and denote the number of such matrices as
\[ N(\bs{r},\bs{c}) := \bigl|\Omega(\bs{r},\bs{c})\bigr| . \]
The ideal goal is to sample from the uniform distribution over $\Omega(\bs{r},\bs{c})$, called the {\em target distribution} and denoted as
\[ P(\bs{z}) := \frac{1}{N(\bs{r},\bs{c})}\ind\bigl\{\bs{z}\in\Omega(\bs{r},\bs{c})\bigr\} \]
where $\ind$ is the indicator function and where $\Omega(\bs{r},\bs{c})$ is assumed to be non-empty.  Unfortunately, existing algorithms for exact uniform sampling are too slow for many applications.  Importance sampling is a potential remedy, requiring the more modest goal of an {\em approximately} uniform distribution $Q$, called the {\em proposal distribution}, that does permit fast sampling and for which $Q(\bs{z})$ can be easily evaluated for any $\bs{z}$.  
This paper describes a strategy for creating such a proposal distribution that often works well in practice.  The techniques here are inspired by the approximately uniform proposal distribution introduced by \cite{Chen:Sequential:2005} (henceforth, CDHL).  The main innovation here is the use of dynamic programming (DP) to combine algebraic constraints with combinatorial approximations.

\section{Introduction}

Binary matrices (zero-one tables, directed graphs, bipartite graphs) arise in a variety of statistical contexts.  They are used, for example, to represent occurrence matrices in ecology, discretized point processes in neurophysiology, and connectivity graphs for social networks.
Random generation of binary matrices is an important computational tool in these settings, especially for conditional inference, in the spirit of Fisher's exact test for independence in two-way contingency tables.  CDHL describe an algorithm for {\em approximate} uniform generation of binary matrices with fixed row and column sums (margins, degree sequences).  The uniform distribution is natural for a variety of conditional inference tests (and also for approximate counting), but exact sampling from the uniform distribution over binary matrices with specified margins is computationally prohibitive.

Although the CDHL distribution is only approximately uniform, the observations can be reweighted to give consistent inferences via {\em importance sampling}.  In Appendix \ref{s:is} we discuss importance sampling as it is used in this context.  See \cite{Liu:Monte:2001} for more details about importance sampling and see \cite{chen2007cit} for further discussion of the importance sampling approach to random generation of matrices for statistical applications.  

The appeal of importance sampling is that the departure from uniformity can be quantified {\em and corrected}, because the probability of generating a given matrix under the proposal distribution is known exactly.
Other approaches for uniform sampling include Markov chain Monte Carlo (MCMC) algorithms with a uniform stationary distribution \citep[e.g.,][]{besag1989gmc, bezakova2007sbc, verhelst2008ema} and randomized construction algorithms for which the probabilities of constructing a given matrix are {\em unknown} but which are asymptotically uniform in the size of the matrix \citep[e.g.,][]{Bayati:Sequential:2009}.  A drawback of MCMC approaches is that the mixing time can be difficult to quantify and will often be prohibitively long for large matrices (especially large sparse matrices).  The MCMC algorithm in \cite{bezakova2007sbc}, for example, mixes in polynomial time, but the polynomial is of a rather high order.  Using an MCMC algorithm before the stationary distribution is reached can lead to convergence failure for Monte Carlo estimates, or worse, apparent convergence to the wrong quantity.  A generic importance sampling algorithm may exhibit similar pathologies \citep[][give an example with the CDHL proposal distribution]{bezakova2006nes}, but the proposal probabilities (the importance weights) tend to provide much more diagnostic information about convergence problems.  Asymptotically uniform randomized construction algorithms provide very little diagnostic information and one can only hope that the asymptotics are valid.  Recently, \cite{blanchet2006isa} has shown that a variant of the CDHL proposal distribution is asymptotically uniform in a certain sense, which represents the best of both worlds: the diagnostic and correction capabilities of importance sampling combined with guarantees that large problems will not create pathological behavior in the importance sampling estimates.  The proof techniques in \cite{blanchet2006isa} should also extend to the proposal distributions described in this paper, but no attempt is made to do so here.

Following CDHL, our goal in this paper is to create a proposal distribution, $Q$, over binary matrices that permits fast Monte Carlo sampling and that is as close as possible to the target distribution $P$, which is uniform subject to specified margins.  As CDHL demonstrated and as we will see below, in situations with sparse or approximately constant margins it is possible to create practical proposal distributions that are extremely close to uniform.  Consequently, there is little or no loss of statistical efficiency compared with exact uniform sampling.  

This paper describes some interesting modifications of the CDHL algorithm based on dynamic programming (DP) and on the recent asymptotic approximations in \cite{canfield2008aed} and \cite{greenhill2006aes}.  Much like the CDHL procedure, the algorithms require $O\bigl(m\sum_{j=1}^n c_j\bigr)$ operations to generate a single sample, which scales well to large, sparse problems.  The modified algorithms perform extremely well in many cases, including cases with irregular margins.  The improved performance over existing approaches is primarily a result of using the improved approximations in \cite{canfield2008aed} and \cite{greenhill2006aes}.  While this improvement is, practically speaking, quite significant, it simply replaces the approximations used in CDHL with more recent findings.  The DP perspective, on the other hand, presents a more natural way to incorporate structural constraints, like fixed margins or forced zeros, into an importance sampling approach.  It substantially simplifies implementation, generalizes immediately to symmetric and integer-valued matrices and seems likely to be broadly applicable.  Of the two modifications introduced here, the DP perspective seems to be the more theoretically interesting.

We leave the investigation of most generalizations to future work and focus here on the situation where the row sums, $\bs{r}$, and column sums, $\bs{c}$, are fixed and the entries are constrained to be either zero or one.  After describing this simple case, we generalize slightly to the case where at most one entry in each row and column is constrained to be zero, which includes the useful case of a zero diagonal.
\cite{chen2007cit} also modified the CDHL procedure to allow for structural zeros.  

\section{An approximately uniform proposal distribution for binary matrices}

This section describes the proposal distribution, $Q$, along with an efficient sampling algorithm.  A key feature is that $Q$ is defined (and sampled) recursively over columns.  Some notational remarks:  We continue with the notation and precise problem statement from Section \ref{s:problem}, including the use of bold font to denote vectors/matrices and normal font (with subscripts) to denote the elements of the corresponding vector/matrix, so that, for example, $\bs{t}:=(t_1,\dotsc,t_k)$.  We also use the notation $:=$ to mean {\em defined as}, which can be important since several quantities are redefined throughout the text as we consider different underlying approximations.  

Exact sampling from $P$ is impractical.  Nevertheless, as CDHL note, in order to sample from $P$ we only need to solve the (generic) problem of sampling from the distribution of the first column, which we denote as 
\begin{equation} P_1(\bs{b}) := \frac{N(\bs{r}-\bs{b},\bs{c'})}{N(\bs{r},\bs{c})}\ind\bigl\{\bs{b}\in\Omega_1(\bs{r},\bs{c})\bigr\} \label{e:P*} \end{equation}
where the support of this distribution is the set of feasible first columns, namely,
\[ \Omega_1(\bs{r},\bs{c}) := \bigl\{\bs{b}\in\{0,1\}^m : \exists \bs{z}\in\Omega(\bs{r},\bs{c}) \text{ with } b_i=z_{i1}, \forall i\bigr\} \]
and where
\[ \bs{c'} := (c_2,\dotsc,c_n) . \]
(We abuse notation throughout and use $\Omega$ and $N$ for matrices of any size, as determined by the lengths of the margin vectors.)
To see that sampling from $P_1$ is enough for the ultimate goal of sampling from $P$, note that the conditional distribution of the remaining sub-matrix given that the first column equals $\bs{b}$ is simply the uniform distribution over $\Omega(\bs{r}-\bs{b},\bs{c'})$.  So we can recursively sample the columns, updating the new margins of the remaining submatrix based on the previously sampled columns.  

Exact sampling from $P_1$ is, of course, still impractical, but it suggests a useful strategy for constructing a proposal distribution $Q$.  Namely, we can create a distribution $Q_1$ that is close to $P_1$ and sample the first column from $Q_1$.  Then we can recursively sample the columns, updating the margins as we go.  This recursive procedure implicitly defines a proposal distribution $Q$ over the entire matrix, but we need only concern ourselves with $Q_1$, the proposal distribution for the first column.  This is often called {\em sequential importance sampling}, since the importance sampling distribution is defined in a sequential manner.

Our goal then is to design a proposal distribution $Q_1$ that is close to $P_1$.
As CDHL observe, approximations to $N(\bs{r},\bs{c})$ can be used to formulate sensible choices for $Q_1$.  In general, if we are given an approximation
\[ \widetilde N(\bs{r},\bs{c}) \approx N(\bs{r},\bs{c})  \]
with the property that $\widetilde N(\bs{r},\bs{c}) > 0$ whenever $N(\bs{r},\bs{c}) > 0$,
then we can try to approximate $P_1(\bs{b})$ in \eqref{e:P*} by
\[ P_1(\bs{b}) \approx \frac{\widetilde N(\bs{r}-\bs{b},\bs{c'})}{\widetilde N(\bs{r},\bs{c})}\ind\bigl\{\bs{b}\in\Omega_1(\bs{r},\bs{c})\bigr\} . \]
Renormalizing this approximation to be a valid probability distribution gives a candidate proposal distribution defined by
\begin{equation} \label{e:Q*} Q_1(\bs{b}) \propto \widetilde N(\bs{r}-\bs{b},\bs{c'}) \ind\bigl\{\bs{b}\in\Omega_1(\bs{r},\bs{c})\bigr\}  \end{equation}
where $\propto$ means ``proportional to'' as a function of $\bs{b}$.

The remarkable fact is that \eqref{e:Q*} permits fast sampling via dynamic programming (DP) when used in conjunction with some of the best known asymptotic approximations for $N(\bs{r},\bs{c})$.  Before giving all of the details in subsequent sections, we give an overview of the main ideas.  To be precise we need to assume that $r_1 \geq r_2 \geq \dotsb \geq r_m$.  This is not a problem since the rows can be permuted and then un-permuted in deterministic preprocessing and postprocessing steps, as discussed in more detail below.   

The first important observation (Section \ref{s:N}) is that for these approximations
\begin{equation} \widetilde N(\bs{r}-\bs{b},\bs{c'}) \propto  \prod_{i=1}^m p_i^{b_i}(1-p_i)^{1-b_i} \label{e:Nf} \end{equation}
for some {\em easily computable} constants $\bs{p}\in[0,1]^m$ depending only on $\bs{r}$ and $\bs{c}$, and for $\bs{b}\in\Omega_1(\bs{r},\bs{c})$.  This means that
\begin{equation} \label{e:f} Q_1(\bs{b}) \propto \prod_{i=1}^m p_i^{b_i}(1-p_i)^{1-b_i} \ind\bigl\{\bs{b}\in\Omega_1(\bs{r},\bs{c})\bigr\} \end{equation}
which is the conditional distribution of a sequence of $m$ independent Bernoulli$(p_i)$ random variables (rv's) given that the sequence is in $\Omega_1(\bs{r},\bs{c})$.  

The second important observation (Section \ref{s:O}) is that the structure of $\Omega_1(\bs{r},\bs{c})$ also factors as
\begin{equation} \label{e:Omega} \ind\bigl\{\bs{b}\in\Omega_1(\bs{r},\bs{c})\bigr\} = \prod_{i=1}^m \ind\{b_i\in A_i\}\ind\bigl\{\textstyle\sum_{\ell=1}^i b_\ell \in B_i\bigr\} \end{equation}
for some {\em easily computable} sets $A_1,\dotsc,A_m \subseteq \{0,1\}$ and $B_1,\dotsc,B_m \subseteq \{0,1,\dotsc,c_1\}$.  This factorization is a consequence of the Gale-Ryser conditions for the existence of a binary matrix with certain specified margins \citep{gale1957tfn,ryser1957cpm}.  Combining the two pieces gives
\begin{equation} \label{e:Qf} Q_1(\bs{b}) \propto \prod_{i=1}^m p_i^{b_i}(1-p_i)^{1-b_i}\ind\{b_i\in A_i\}\ind\bigl\{\textstyle\sum_{\ell=1}^i b_\ell \in B_i\bigr\} \end{equation}

The third and final important observation (Section \ref{s:DP}) is that \eqref{e:Qf} defines a Markov chain on the partial sums of the first column vector.  In particular, using the change of variables
\[ s_i = \sum_{\ell=1}^i b_\ell \quad \quad i = 1,\dotsc,m \]
and defining $s_0=0$ for convenience, we have that
\begin{equation} \label{e:Qs} Q_1(\bs{s}) \propto \prod_{i=1}^m p_i^{s_i-s_{i-1}}(1-p_i)^{1-(s_i-s_{i-1})}\ind\{s_i-s_{i-1}\in A_i\}\ind\bigl\{s_i \in B_i\bigr\} \end{equation}
Since $Q_1(\bs{s})$ factors into terms that only involve consecutive $(s_{i-1},s_i)$ pairs, it must be a Markov chain.  DP can be used to efficiently recover the transition probability matrices, at which point sampling the sequence of partial sums, and hence the first column, is trivial.  
In the next three sections (Sections \ref{s:N}--\ref{s:DP}) we provide details for each of these three main observations and then in Section \ref{s:overview} we give an overview of the entire sampling algorithm.

CDHL essentially noted both \eqref{e:f} and \eqref{e:Omega}, but they did not use DP to combine the two.  Instead, they tried two different approaches.  The first began with \eqref{e:f} but replaced $\Omega_1(\bs{r},\bs{c})$ with the larger set $\{\bs{b}:\sum_i b_i = c_1\}$, namely, that the only constraint is that the total number of ones in the first column is correct.  This simplified constraint reduces \eqref{e:f} to the case of conditional Bernoulli sampling for which efficient sampling algorithms exist \citep{chen1997sap}.  (Interestingly, DP is one such algorithm, but not the one used by CDHL.)  The simplified constraint makes the support of $Q_1$ larger than the support of $P_1$ (in other words, the proposal generates invalid matrices) and can substantially reduce efficiency.  To remedy this, CDHL make use of the factorization in \eqref{e:Omega}, but they do not directly combine it with \eqref{e:f}.  Instead, they use a heuristic procedure in which the proposal distribution becomes a complicated mixture of distributions inspired by \eqref{e:f} and designed to exactly cover the support of $P$.  We have not investigated the degree to which the heuristic procedure in CDHL agrees with the DP approach here.  Certainly, the DP approach is much easier to implement and to analyze.  The computational costs are more or less identical. 

\subsection{Approximations to $N(r,c)$} \label{s:N}

The goal is to find an approximation to $N(\bs{r},\bs{c})$ that permits the factorization in \eqref{e:Nf}.    
Here is a simple, but generic result along these lines.
\begin{lem} \label{l:f}
Suppose that
\begin{equation} \widetilde N(\bs{r},\bs{c}) =  g_0(\bs{c},m,n)\prod_{i=1}^{m} g_i(r_i,\bs{c},m,n) \label{e:lemf} \end{equation}
whenever $N(\bs{r},\bs{c}) > 0$ for some functions $(g_i:i=0,\dotsc,m)$, where $m$ and $n$ are the respective number of elements of $\bs{r}$ and $\bs{c}$.  Then as $\bs{b}$ varies over $\Omega_1(\bs{r},\bs{c})$, \eqref{e:Nf} holds with 
\begin{equation} \label{e:pi} p_i := \frac{g_i(r_i-1,\bs{c'},m,n-1)}{g_i(r_i,\bs{c'},m,n-1)+g_i(r_i-1,\bs{c'},m,n-1)} . \end{equation}
\end{lem}

Lemma \ref{l:f} applies to most of the approximations to $N(\bs{r},\bs{c})$ that have appeared in the literature, but not to all of them.  If Lemma \ref{l:f} is not applicable, then the factorization in \eqref{e:Nf} is unlikely to hold exactly.  Nevertheless, we can force the factorization in \eqref{e:Nf} with the following heuristic based on a Taylor's approximation.  Define $\bs{1^i}$ to be the $m$-length binary vector with a one at entry $i$ and zeros elsewhere.  Then
\begin{align}
& \widetilde N(\bs{r}-\bs{b},\bs{c'}) = \exp\bigl(\log \widetilde N(\bs{r}-\bs{b},\bs{c'})\bigr) 
\notag \\ & \quad \approx \exp\left(\log \widetilde N(\bs{r},\bs{c'}) + \textstyle\sum_{i=1}^m \bigl(\frac{\partial}{\partial r_i} \log \widetilde N(\bs{r},\bs{c'})\bigr)(r_i-b_i-r_i)\right)
\notag \\ & \quad \approx \widetilde N(\bs{r},\bs{c'}) \exp\left(-\textstyle\sum_{i=1}^m b_i \bigl(\log \widetilde N(\bs{r},\bs{c'})-\log \widetilde N(\bs{r}-\bs{1^i},\bs{c'})\bigr)\right)
\notag \\ & \quad \propto \prod_{i=1}^m p_i^{b_i}(1-p_i)^{1-b_i} \notag
\notag \end{align}
for $\bs{p}$ defined by
\begin{equation} \label{e:pi2} p_i := \frac{\widetilde N(\bs{r}-\bs{1^i},\bs{c'})}{\widetilde N(\bs{r},\bs{c'}) + \widetilde N(\bs{r}-\bs{1^i},\bs{c'})} . \end{equation}
\eqref{e:pi} and \eqref{e:pi2} give the same values for $\bs{p}$ whenever Lemma \ref{l:f} applies, but \eqref{e:pi2} can be applied to any approximation, as long as it extends nicely to the potentially invalid margins $(\bs{r},\bs{c'})$ and $(\bs{r}-\bs{1^i},\bs{c'})$.  Note, however, that if Lemma \ref{l:f} does not apply, then using $\bs{p}$ defined by \eqref{e:pi2} will lead to a proposal distribution that is different from \eqref{e:Q*} in that the original approximation, $\widetilde N$, has been replaced by an additional Taylor's approximation.    

We suggest two approximations to $N(\bs{r},\bs{c})$ that have been developed recently.  Each of these are refinements of the suggestions in CDHL.  For each positive integer $\ell$ and any nonnegative integer $a$ we define 
\[ [a]_\ell := a(a-1)\dotsm(a-\ell+1) . \]
For a $k$-vector $\bs{t}$ of nonnegative integers we define
\[ \textstyle [\bs{t}]_\ell := \sum_{i=1}^k [t_i]_\ell . \]
Note that $[\bs{r}]_1=[\bs{c}]_1$ whenever $N(\bs{r},\bs{c}) > 0$, since each gives the number of ones in any binary matrix compatible with these margins.

\subsubsection{Canfield, Greenhill \& McKay (2008)}

The first approximation that we consider comes from \cite{canfield2008aed} and is accurate (asymptotically) as long as the margins do not vary too wildly:
\begin{gather} \label{e:n1}
\widetilde N(\bs{r},\bs{c}) := \binom{mn}{[\bs{c}]_1}^{-1}\prod_{i=1}^m \binom{n}{r_i}\prod_{j=1}^n \binom{m}{c_j} \exp\left(-\frac{1}{2}\bigl(1-\mu\bigr)\bigl(1-\nu\bigr)\right) \\
\mu := \mu(\bs{r},\bs{c},m,n) := \frac{mn}{[\bs{c}]_1\bigl(mn-[\bs{c}]_1\bigr)}\sum_{i=1}^m \bigl(r_i-[\bs{c}]_1/m\bigr)^2  \notag \\
\nu := \nu(\bs{c},m,n) := \frac{mn}{[\bs{c}]_1\bigl(mn-[\bs{c}]_1\bigr)}\sum_{j=1}^n \bigl(c_j-[\bs{c}]_1/n\bigr)^2 \notag .
\end{gather}
\eqref{e:n1} permits the factorization in Lemma \ref{l:f} with
\begin{subequations} 
\[ g_i(r_i,\bs{c},m,n) := \binom{n}{r_i}\exp\left(\frac{mn\bigl(r_i-[\bs{c}]_1/m\bigr)^2}{2[\bs{c}]_1\bigl(mn-[\bs{c}]_1\bigr)}\bigl(1-\nu(\bs{c},m,n)\bigr)\right) . \] 
\eqref{e:pi} gives the Bernoulli probabilities, which simplify to
\begin{gather} p_i := \frac{r_i\exp\Bigl(\beta\bigl(1-2(r_i-[\bs{c'}]_1/m)\bigr)\Bigr)}{n-r_i+r_i\exp\Bigl(\beta\bigl(1-2(r_i-[\bs{c'}]_1/m)\bigr)\Bigr)}  \label{e:g1} 
\\ \notag \beta := \frac{m(n-1)}{2[\bs{c'}]_1\bigl(m(n-1)-[\bs{c'}]_1\bigr)}\bigl(1-\nu(\bs{c'},m,n-1)\bigr)  \end{gather}
where we take $0/0:=0$.
Note that \eqref{e:n1} is an improvement over the first approximation in CDHL which was just \eqref{e:n1} without the exponential term, giving $g_i(r_i,\bs{c},m,n) := \binom{n}{r_i}$
and leading to the Bernoulli probabilities 
\begin{equation}   \label{e:g1old} p_i :=\frac{r_i}{n} .\end{equation}
\end{subequations} 
In most examples \eqref{e:g1} is a substantial improvement over \eqref{e:g1old}.  \eqref{e:g1} tends to work well as long as the margins do not vary wildly.  For certain pathological cases with wildly varying margins \cite{bezakova2006nes} have shown that \eqref{e:g1old} fails completely.  \eqref{e:g1} also fails on these examples. 

\subsubsection{Greenhill, McKay \& Wang (2006)}

The second approximation comes from \cite{greenhill2006aes} and is accurate (asymptotically) for sparse matrices, except perhaps when the margins are extremely variable:
\begin{gather} \label{e:n2}
\widetilde N(\bs{r},\bs{c}) := \frac{[\bs{c}]_1!}{\prod_{i=1}^m r_i!\prod_{j=1}^nc_j!}\exp\bigl(-\alpha_1(\bs{c})[\bs{r}]_2-\alpha_2(\bs{c})[\bs{r}]_3-\alpha_3(\bs{c})[\bs{r}]_2^2\bigr) \\
\begin{aligned}
\alpha_1(\bs{c}) & := \frac{[\bs{c}]_2}{2[\bs{c}]_1^2}+\frac{[\bs{c}]_2}{2[\bs{c}]_1^3}+\frac{[\bs{c}]_2^2}{4[\bs{c}]_1^4} \notag \\
\alpha_2(\bs{c}) & := -\frac{[\bs{c}]_3}{3[\bs{c}]_1^3}+\frac{[\bs{c}]_2^2}{2[\bs{c}]_1^4} \notag \\
\alpha_3(\bs{c}) & := \frac{[\bs{c}]_2}{4[\bs{c}]_1^4}+\frac{[\bs{c}]_3}{2[\bs{c}]_1^4}-\frac{[\bs{c}]_2^2}{2[\bs{c}]_1^5} \notag
\end{aligned} \notag
\end{gather}
where we take $0/0:=0$.
Lemma \ref{l:f} does not apply because the $[\bs{r}]_2^2$ term does not factor appropriately.  Using the additional approximation in \eqref{e:pi2} leads to 
\begin{subequations}
\begin{gather} \label{e:g2}
p_i := \frac{r_i\exp\bigl(\gamma(r_i-1)\bigr)}{1+r_i\exp\bigl(\gamma(r_i-1)\bigr)} \\ \notag
\gamma := 2\alpha_1(\bs{c'})+3\alpha_2(\bs{c'})(r_i-2)+4\alpha_3(\bs{c'})\bigl([\bs{r}]_2-r_i+1\bigr)
\end{gather}
\eqref{e:n2} is an improved version of the approximation in \cite{Oneil:Asymptotics:1969}, which was mentioned in CDHL, although they did not use it for any of their reported results.  The \cite{Oneil:Asymptotics:1969} approximation replaces the argument of the exponential in \eqref{e:n2} with $-[\bs{r}]_2[\bs{c}]_2/\bigl(2[\bs{c}]_1^2\bigr)$.  Lemma \ref{l:f} now applies with  
\[ g_i(r_i,\bs{c},m,m) := \frac{1}{r_i!}\exp\left(-\frac{[r_i]_2[\bs{c}]_2}{2[\bs{c}]_1^2}\right) \]
leading to 
\begin{equation} \label{e:g2old}
p_i := \frac{r_i\exp\bigl((r_i-1)[\bs{c'}]_2/[\bs{c'}]_1^2\bigr)}{1+r_i\exp\bigl((r_i-1)[\bs{c'}]_2/[\bs{c'}]_1^2\bigr)}
\end{equation}
where we take $0/0:=0$.
\end{subequations}
\eqref{e:g2} works extremely well as long as the resulting binary matrices are sparse, but tends to break down quickly as the matrices become more dense.  \cite{blanchet2006isa} has shown that \eqref{e:g2old} leads to sampling procedures that are exponentially efficient for large, sparse matrices.  Presumably, \eqref{e:g2} behaves at least as well.  Both happen to be exactly uniform for the pathological cases in \cite{bezakova2006nes}.

\subsection{The structure of $\Omega_1(r,c)$} \label{s:O}

In this section we give the details of the factorization in \eqref{e:Omega}.

For any $k$-vector $\bs{t}$ we define $t^*_j := \#\{i:t_i \geq j\}$ to be the number of elements of $\bs{t}$ that are greater than or equal to $j$, where $j=1,2,\dotsc$.  The sequence $\bs{t^*}:=(t_1^*,t_2^*,\dotsc)$ is called the {\em conjugate} of $\bs{t}$.  The next theorem refers to $\bs{c'^*}$ which is the conjugate of the $(n-1)$-vector $\bs{c'}:=(c_2,\dotsc,c_n)$.  

\begin{thm} \cite{Chen:Sequential:2005}. \label{t} Assume that $r_1\geq r_2 \geq\dotsb\geq r_m$
and that $N(\bs{r},\bs{c}) > 0$.  Define $\bs{A}:=A_1\times\dotsb\times A_m$ and $\bs{B}:=B_1\times\dotsb\times B_m$ by
\begin{align*} A_i & := \begin{cases}  \{0\} & \text{if $r_i=0$;}
 \\ \{0,1\} & \text{if $0 < r_i < n$;} \\
 \{1\} & \text{if $r_i=n$;} \end{cases} \\
B_i & := \begin{cases} \textstyle \Bigl\{\max\bigl\{0,\sum_{\ell=1}^i r_\ell-\textstyle \sum_{\ell=1}^i c'^*_\ell\bigr\} , \dotsc, c_1\Bigr\} & \text{if  $i < m$;} \\
\{c_1\} & \text{if $i = m$.} \end{cases}
\end{align*}
Let $\bs{b}$ be a binary $m$-vector and let $\bs{s}$ denote the partial sums of $\bs{b}$ defined by $s_i:=\sum_{\ell=1}^i b_\ell$.   Then $\bs{b}\in\Omega_{1}(\bs{r},\bs{c})$ if and only if $\bs{b}\in \bs{A}$ and $\bs{s}\in \bs{B}$.  
\end{thm}

The factorization in \eqref{e:Omega} is just a restatement of this theorem's conclusions.  It is instructive to see how the theorem arises.  For $r_1\geq\dotsb\geq r_m$, the Gale-Ryser conditions \citep{gale1957tfn,ryser1957cpm} state that $\Omega(\bs{r},\bs{c})$ is non-empty if and only if 
\[ \textstyle \sum_{\ell=1}^i r_\ell \leq \sum_{\ell=1}^i c^*_\ell \quad \text{for $i < m$} \quad \text{and} \quad \sum_{\ell=1}^m r_\ell = \sum_{\ell=1}^m c^*_\ell \]
So a potential first column vector $\bs{b}$ will be valid as long as $\bs{b}\in\bs{A}$ (obvious) and $\Omega(\bs{r}-\bs{b},\bs{c'})$ is not empty (there must be some way to fill the remaining matrix).  Ignoring for the moment that the entries of $\bs{r}-\bs{b}$ may not be decreasing, we can apply the Gale-Ryser conditions to this last constraint to get
\[ \textstyle \sum_{\ell=1}^i (r_\ell-b_\ell) \leq \sum_{\ell=1}^i c'^*_\ell \quad \text{for $i < m$} \quad \text{and} \quad \sum_{\ell=1}^m (r_\ell-b_\ell) = \sum_{\ell=1}^m c'^*_\ell \]
which can be rearranged as
\begin{equation} \textstyle \sum_{\ell=1}^i b_\ell \geq \sum_{\ell=1}^i r_\ell - \sum_{\ell=1}^i c'^*_\ell \quad \text{for $i < m$} \quad \text{and} \quad \sum_{\ell=1}^m b_\ell = c_1  \label{e:GR} \end{equation}
where in the last equality we have used the easy to verify property that $\sum_{\ell=1}^m r_\ell - \sum_{\ell=1}^m c'^*_\ell = c_1$.  Noting that the partial sums must be in $\{0,\dotsc,c_1\}$, we see that \eqref{e:GR} is just the constraint that $\bs{s}\in\bs{B}$.  CDHL prove that nothing goes wrong by ignoring the fact that $\bs{r}-\bs{b}$ may not be decreasing.

\subsection{Dynamic programming} \label{s:DP}

We have now arrived at the factorization of $Q_1$ given in \eqref{e:Qf}.  The goal in this section is to show how DP can be used to efficiently sample from $Q_1$.  The first step is to change variables from the binary vector $\bs{b}$ to its partial sums $\bs{s}$.  The distribution $Q_1$ in \eqref{e:Qf} on $\bs{b}$ is equivalent to the distribution $Q_1$ in \eqref{e:Qs} on $\bs{s}$, which we reproduce here for easy reference:
\[ Q_1(\bs{s}) \propto \prod_{i=1}^m p_i^{s_i-s_{i-1}}(1-p_i)^{1-(s_i-s_{i-1})}\ind\{s_i-s_{i-1}\in A_i\}\ind\bigl\{s_i \in B_i\bigr\} \]
(Recall that we define $s_0=0$ for convenience.  Also note that using $Q_1$ here is an abuse of notation --- we should technically give the distribution over partial sums a new name.)  We can rewrite $Q_1(\bs{s})$ as
\begin{equation} Q_1(\bs{s}) \propto \prod_{i=1}^m h_i(s_{i-1},s_i) \label{e:Qh} \end{equation}
where
\[ h_i(s_{i-1},s_i) := p_i^{s_i-s_{i-1}}(1-p_i)^{1-(s_i-s_{i-1})}\ind\{s_i-s_{i-1}\in A_i\}\ind\bigl\{s_i \in B_i\bigr\} \]
Recall that the constants $\bs{p}$ and the sets $\bs{A}$ and $\bs{B}$ each depend on the margins $\bs{r}$ and $\bs{c}$, and that they are easy to compute and represent as shown in Sections \ref{s:N} and \ref{s:O} above.  So the functions $h_i$ are also easy to compute and represent.  They can always be expressed as $(c_1+1)\times(c_1+1)$ matrices, but this is rather inefficient since, for example, we know that $h_i(s_{i-1},s_i) = 0$ if $s_{i}-s_{i-1}\not\in\{0,1\}$.  A more efficient representation would be as $(c_1+1)\times 2$ matrices, where the rows index the possible values of $s_{i-1}$ and the columns index the two possibilities of $s_i=s_{i-1}$ and $s_i=s_{i-1}+1$.

Let $\bs{S}:=(S_1,\dotsc,S_m)$ denote the random sequence of partial sums with distribution given by \eqref{e:Qh} and define $S_0:=0$ for convenience.  If we were given the standard Markov chain representation
\begin{equation} Q_1(\bs{s}) = \prod_{i=1}^m \pi_i(s_i|s_{i-1}) \label{e:Qpi} \end{equation}
where $\pi_i(s_i|s_{i-1}) := \Prob\bigl(S_i=s_i\bigl|S_{i-1}=s_{i-1}\bigr)$, then generating a random observation of $\bs{S}$ would be trivial.  It is well known that DP can be used to convert from product representations like \eqref{e:Qh} into conditional probability representations like \eqref{e:Qpi}.  The next theorem summarizes DP in this context.

\begin{thm} (See e.g.~\cite{Frey:Graphical:1998}.) \label{t:DP} Let $(S_0,S_1,\dotsc,S_m)$ be a sequence of rv's where each $S_i$ takes values in the finite set $D_i$ and where $D_0:=\{0\}$.  Suppose there exists a sequence of functions $h_i:D_{i-1}\times D_i\mapsto[0,\infty)$ for $i=1,\dotsc,m$ such that the distribution of $(S_1,\dotsc,S_m)$ can be expressed as
\[ \textstyle \Prob\bigl(S_1=s_1,\dotsc,S_m=s_m\bigr) \propto \prod_{i=1}^m h_i(s_{i-1},s_i) \]   Recursively (in reverse) define 
\[ \begin{aligned} \beta_m(s_{m-1},s_m) & := h_m(s_{m-1},s_m) \\
 \beta_i(s_{i-1},s_i) & := h_i(s_{i-1},s_i)\textstyle\sum_{s_{i+1}\in D_{i+1}} \beta_{i+1}(s_i,s_{i+1}) \quad \text{for $i=1,\dotsc,m-1$}, \end{aligned}\]
where each $\beta_i$ is defined over $D_{i-1}\times D_i$.  Then 
\[ \Prob\bigl(S_i=s_i\bigl|S_{i-1}=s_{i-1}\bigr) = \frac{\beta_i(s_{i-1},s_i)}{\sum_{s\in D_{i}} \beta_i(s_{i-1},s)} \] for each $i=1,\dotsc,m$.
\end{thm}

Applying this theorem to the special structure of our problem, we first note that $\beta_i(s_{i-1},s_i) = 0$ whenever $h_i(s_{i-1},s_i)=0$, so, as mentioned earlier, we can also represent each $\beta_i$ as a $(c_1+1)\times 2$ matrix.  Similarly, the sums over $D_i$ and $D_{i+1}$ in the DP algorithm include at most 2 nonzero elements.  This means that the entire DP computation (in our case) takes at most $O(mc_1)$ operations.\footnote{It also important to note that the $\beta_i$'s in the recursion can be scaled by an arbitrary constant (even during the recursion), which is often necessary in practice for controlling underflow and overflow.  A good generic choice is to divide each $\beta_i$ by its sum over both arguments.  This does not affect the order of the number of operations.} 

The result of the DP computation is the standard Markov chain representation for $Q_1(\bs{s})$ in \eqref{e:Qpi}.  Initializing $S_0=0$, we can recursively generate a random observation $\bs{S}$ and its probability $Q_1(\bs{S})$ in at most $O(m)$ operations, including at most $m$ calls to a random number generator.  Then we can convert from the partial sum representation back to the binary vector representation with $m$ subtractions.  This gives a single sample from $Q_1(\bs{b})$.  The entire process of DP and sampling takes $O(mc_1)$ operations.

\subsection{Overview of the algorithm} \label{s:overview}

In this section we bring everything together and present an overview of the algorithm.  Here is the procedure to sample the first column.
\begin{itemize}
\item Reorder the rows so that the row sums are decreasing: $r_1 \geq \dotsb \geq r_m$.
\item Compute the Bernoulli probabilities $\bs{p}$.  Section \ref{s:N} gives some suggestions.
\item Use Theorem \ref{t} to compute the sets $\bs{A}$ and $\bs{B}$.   
\item Use Theorem \ref{t:DP} to compute the Markov chain representation of the partial sums.
\item Generate a random observation of the partial sums from this Markov chain representation and also compute its probability.
\item Deterministically convert the partial sums into the equivalent binary vector.  This is the first column for the reordered rows.
\item Return the rows to their original order.  
\end{itemize}  
Neglecting the computation of $\bs{p}$, the whole process takes $O(mc_1)$ operations.\footnote{Technically, there are an additional $O(m\log m)$ operations required to sort the rows.  However, this need only be done once.  The ordering of the rows can be efficiently updated without generic sorting during the recursive computation to sample the entire matrix.  Furthermore, usually multiple random samples of the matrix are required, which also does not require additional sorting, so, in an efficient implementation, the sorting is best viewed as a negligible preprocessing step.}  The heuristics for choosing $\bs{p}$ described in Section \ref{s:N} require negligible amounts of additional computation.\footnote{Before sampling each column, one can also deterministically assign the remaining entries of any rows and columns whose remaining entries must be all zeros or all ones (as evidenced by trivial row and column sums), and then remove these rows and columns from further consideration (cf., CDHL). This can be viewed as a modification of the heuristics for choosing $\bs{p}$; we do not use it here.}

Once the first column has been sampled, the row and column sums are updated and the process is repeated, recursively, to generate successive columns.  The probabilities of each column are multiplied together to generate the probability of the entire matrix.  Note that the successive values of $\bs{p}$, $\bs{A}$ and $\bs{B}$ will depend adaptively on the previously sampled columns.  Generating an entire $m\times n$ binary matrix takes $O(m[\bs{c}]_1)$ operations, where $[\bs{c}]_1:=\sum_{j=1}^n c_j$ is the total number of ones.  

The degree to which $Q$ approximates $P$ can be affected by a variety of deterministic preprocessing steps, such as, reordering the columns, swapping the roles of ones and zeros, or swapping the roles of rows and columns.  Following CDHL, in the examples below we order the columns in order of decreasing column sums.  (This reordering becomes necessary when we generalize to structural zeros.)  We have not systematically explored preprocessing strategies.  For any given problem, of course, one could explore various types of preprocessing and also various heuristics for choosing $\bs{p}$ in order to select the best proposal distribution.

The identical algorithm can also be used to evaluate $Q(\bs{z})$ for any given $\bs{z}$.  
For each column we simply omit the step of generating a random observation of the partial sums from the Markov chain representation and instead use the partial sums from the corresponding column in $\bs{z}$.  Using the Markov chain representation to compute probabilities remains the same.  We make use of this in some of the examples below.

\section{Examples}

The algorithm was implemented in Matlab and computations carried out on a MacBook laptop with 2 GB of RAM and a 2.16 GHz dual core processor.  The Matlab implementation is available from the author.  The implementation can handle moderately large matrices, if they are sparse, although very large matrices of the type encountered with connectivity graphs in large social networks are still out of reach for a laptop.  For example, generating a single observation for a $10^5\times 10^5$ matrix with all row and columns sums equal to 2 takes about 2.5 hours, whereas generating a single such $10^3\times 10^3$ matrix takes about a second.

Let $\bs{Z_1},\dotsc,\bs{Z_N}$ be iid observations (each a binary matrix) from the proposal distribution $Q$ over $\Omega(\bs{r},\bs{c})$ and let $W_1,\dotsc,W_N$ be the (unnormalized) importance weights defined by $W_k := Q\bigl(\bs{Z_k}\bigr)^{-1}$.
We quantify the empirical performance of an importance sampling procedure with two summary statistics based on the importance weights.  The first is the ratio of maximum to minimum importance weights and the second is an estimate of the coefficient of variation, namely, 
\[ \widehat\Delta := \frac{\max_{k=1,\dotsc,N} W_k}{\min_{k=1,\dotsc,N} W_k} \quad \quad\ \text{and} \quad \quad \widehat{cv}{}^2 := \frac{S_W^2}{\overline W^2} \]
where $\overline W$ and $S_W^2$ are the sample mean and sample variance, respectively, of $W_1,\dotsc,W_N$.  We consider $\widehat\Delta\approx 1$ and $\widehat{cv}{}^2 \approx 0$ to be indicative of good performance, although it is easy to imagine situations where these would be misleading.  We also report $\overline W$ and its estimated standard deviation, $S_{\overline W}:=\sqrt{S_W^2/n}$.  Note that $\overline W$ is a consistent point estimate of $N(\bs{r},\bs{c})$.  See CDHL for more details about using and evaluating importance sampling procedures in this context.

Table \ref{t:time} details the speed of the algorithm on $1000\times 1000$ binary matrices with all row and column sums identical.  There is no practical difference in speed between the different heuristics for choosing $\bs{p}$.  These run-times are merely meant to provide a feel for how the algorithm behaves --- no attempt was made to control the other processes operating simultaneously on the laptop.  Presumably a careful C or assembly language implementation would run much faster.  
\begin{table}[h]
\caption{Time needed to generate a single observation using $m=n=1000$ and $r_1=r_i=c_j$.}
\begin{tabular}{|c||c|c|c|c|c|c|c|c|c|} \hline
$r_1$ & 2 & 4 & 8 & 16 & 32 & 64 & 128 & 256 & 512 \\ \hline \hline
time$/N$ & 1.2 s & 1.6 s & 2.4 s & 4.0 s & 6.7 s & 12.4 s & 24.4 s & 39.2 s & 46.6 s \\ \hline
\end{tabular}
\label{t:time}
\end{table}

Table \ref{t:WN1_1} reports $\widehat\Delta$ on these examples using $N=1000$ for each of the heuristics \eqref{e:g1}, \eqref{e:g1old}, \eqref{e:g2}, and \eqref{e:g2old} for choosing $\bs{p}$.
(In the tables, $\infty$ means that the floating point precision of the machine was exceeded.)  \eqref{e:g1} was the best in each case --- the maximum and minimum importance weight are within a few percent of each other --- and it also had the smallest $\widehat{cv}{}^2$ in each case.  Table \ref{t:cv_1} reports all of our measures of importance sampling performance on these examples for the best case of \eqref{e:g1}.

\begin{table}[h]
\caption{$\widehat\Delta$ for $N=1000$ using $m=n=1000$ and $r_1=r_i=c_j$ for all $i,j$}
\begin{tabular}{|c||c|c|c|c|c|c|c|c|c|} \hline
$r_1$ & 2 &   4 &   8 &  16 &  32 &  64 & 128 & 256 & 512  \\ \hline \hline
\eqref{e:g1} & 1.049 &   1.075 &   1.041 &   1.008 &   1.005 &   1.004 &   1.004 &   1.004 &   1.004  \\ \hline
\eqref{e:g1old} & 1.390 &   1.618 &   1.493 &   1.449 &   1.585 &   1.486 &   1.621 &   1.512 &   1.574 \\ \hline
\eqref{e:g2} & 1.080 &   1.082 &   1.146 &   1.375 &   2.610 &   4.769 &     $\infty$ &     $\infty$  &     $\infty$ \\ \hline
\eqref{e:g2old} & 1.172 &   1.215 &   1.487 &   2.216 &   4.788 & 162.9 &     $\infty$ &     $\infty$ &     $\infty$ \\ \hline
\end{tabular}
\label{t:WN1_1}
\end{table}

\begin{table}[h]
\caption{Performance of \eqref{e:g1} for $N=1000$ using $m=n=1000$ and $r_1=r_i=c_j$ for all $i,j$}
\begin{tabular}{|c||c|c|c|} \hline
$r_1$$\vphantom{\hat{\bigl(}}$ & $\widehat\Delta$ & $\widehat{cv}{}^2$ & $\overline W \pm S_{\overline W}$  \\ [1ex] \hline \hline
2 $\vphantom{\bigl(}$ & 1.049 & $4.2\times 10^{-6}$ & $(1.75148 \pm 0.00011) \times 10^{5133}$ \\ \hline
4 $\vphantom{\bigl(}$ & 1.075 & $6.4\times 10^{-6}$ & $(7.64296 \pm 0.00061) \times 10^{9910}$ \\ \hline
8 $\vphantom{\bigl(}$ & 1.041 & $2.1\times 10^{-6}$ & $(1.01879 \pm 0.00005) \times 10^{18531}$ \\ \hline
16 $\vphantom{\bigl(}$ & 1.008 & $3.9\times 10^{-7}$ & $(2.31580 \pm 0.00005) \times 10^{33629}$ \\ \hline
32 $\vphantom{\bigl(}$ & 1.005 & $2.3\times 10^{-7}$ & $(6.50167 \pm 0.00010) \times 10^{59218}$ \\ \hline
64 $\vphantom{\bigl(}$ & 1.004 & $2.2\times 10^{-7}$ & $(1.22048 \pm 0.00002) \times 10^{100716}$ \\ \hline
128 $\vphantom{\bigl(}$ & 1.004 & $1.8\times 10^{-7}$ & $(9.38861 \pm 0.00013) \times 10^{163302}$ \\ \hline
256 $\vphantom{\bigl(}$ & 1.004 & $2.2\times 10^{-7}$ & $(6.70630 \pm 0.00010) \times 10^{243964}$ \\ \hline
512 $\vphantom{\bigl(}$ & 1.004 & $2.2\times 10^{-7}$ & $(5.02208 \pm 0.00007) \times 10^{297711}$ \\ \hline
\end{tabular}
\label{t:cv_1}
\end{table}
 
For some less balanced $50\times 100$ examples, take $\bs{\tilde r}:= $ (24, 22, 22, 17, 17, 17, 17, 13, 13, 13, 12, 12, 11, 11, 11, 10, 10, 9, 9, 9, 8, 8, 8, 8, 8, 8, 7, 6, 6, 6, 6, 5, 5, 5, 5, 4, 4, 4, 4, 4, 3, 3, 3, 3, 3, 3, 2, 2, 2, 2) and $\bs{\tilde c}:=$ (12, 12, 10, 10, 9, 9, 9, 9, 9, 8, 8, 8, 8, 7, 7, 7, 7, 7, 7, 6, 6, 6, 6, 6, 6, 6, 6, 6, 6, 6, 5, 5, 5, 5, 5, 5, 5, 5, 5, 5, 4, 4, 4, 4, 4, 4, 4, 4, 4, 4, 4, 4, 4, 4, 4, 4, 4, 4, 3, 3, 3, 3, 3, 3, 3, 3, 3, 2, 2, 2, 2, 2, 2, 2, 2, 2, 2, 2, 2, 2, 1, 1, 1, 1, 1, 1, 1, 1, 1, 1, 1, 1, 1, 1, 1, 1, 1, 1, 1, 1) and consider row and column sums of the form $\bs{r}:=k\bs{\tilde r}$ and $\bs{c}:=k\bs{\tilde c}$ for $k=1,\dotsc,4$.  Generating a single observation takes between 0.012 s and 0.016 s depending on $k$.  Table \ref{t:WN1_2} summarizes $\widehat\Delta$ as before, but this time with $N=10^5$.  \eqref{e:g1} is best except for the case $k=1$, when \eqref{e:g2} is more uniform.  In general, the performance is much worse than in the regular case and using any of these importance sampling algorithms for the cases $k=3,4$ seems suspicious because the proposal distributions are clearly very far from uniform.  Table \ref{t:cv_2} reports all of our measures of importance sampling performance for \eqref{e:g1}.  For $k=1$, \eqref{e:g2} gives $\widehat{cv}{}^2 = 4.4\times 10^{-4}$ and $\overline W \pm S_{\overline W} = (2.3071\pm .0002)\times 10^{444}$.

\begin{table}[h]
\caption{$\widehat\Delta$ for $N=10^5$ using $\bs{r}=k\bs{\tilde r}$ and $\bs{c}=k\bs{\tilde c}$.}
\label{t:WN1_2}
\begin{tabular}{|c||c|c|c|c|} \hline
$k$ & 1 & 2 & 3 & 4  \\ \hline \hline
\eqref{e:g1} $\vphantom{\bigl(}$  & 3.248 & 6.825 & 299.7 & $4.19\times 10^{8}$ \\ \hline
\eqref{e:g1old} $\vphantom{\bigl(}$  & 145.1 & 918.6 & $5.09\times 10^{7}$ & $9.47\times 10^{13}$ \\ \hline
\eqref{e:g2} $\vphantom{\bigl(}$  & 1.225 & 19.86 & $2.74\times 10^{5}$ & $7.66\times 10^{17}$ \\ \hline
\eqref{e:g2old} $\vphantom{\bigl(}$  & 3.823 & $7.33\times 10^{3}$ & $3.97\times 10^{12}$ & $2.68\times 10^{33}$ \\ \hline
\end{tabular}
\end{table}

\begin{table}[h]
\caption{Performance of \eqref{e:g1} for $N=10^5$ using $\bs{r}=k\bs{\tilde r}$ and $\bs{c}=k\bs{\tilde c}$.}
\begin{tabular}{|c||c|c|c|} \hline
$k$$\vphantom{\hat{\bigl(}}$ & $\widehat\Delta$ & $\widehat{cv}{}^2$ & $\overline W \pm S_{\overline W}$  \\ [1ex] \hline \hline
1 $\vphantom{\bigl(}$ & 3.248 & $1.9\times 10^{-3}$ & $(2.3069 \pm 0.0003) \times 10^{444}$ \\ \hline
2 $\vphantom{\bigl(}$ & 6.825 & $3.3\times 10^{-2}$ & $(2.7975 \pm 0.0016) \times 10^{678}$ \\ \hline
3 $\vphantom{\bigl(}$ & 299.671 & $5.4\times 10^{-1}$ & $(1.9741 \pm 0.0046) \times 10^{766}$ \\ \hline
4 $\vphantom{\bigl(}$ & $4.19\times 10^8$ & $2.8\times 10^{1}$ & $(8.8545 \pm 0.1482) \times 10^{691}$ \\ \hline
\end{tabular}
\label{t:cv_2}
\end{table}

\cite{bezakova2006nes} investigates the performance of CDHL's algorithm on pathological margins with very large $r_1$ and $c_1$, but with all other row and column sums exactly $1$.  They prove that \eqref{e:g1old} is extremely far from uniform for such margins.  Empirically speaking, it seems that \eqref{e:g1} is also quite far from uniform in such cases.  Conversely, it is straightforward to show that \eqref{e:g2} and \eqref{e:g2old} are exactly uniform for these types of margins.  Following \cite{bezakova2006nes}, we experiment with the margins $\bs{r}=(240,1,\dotsc,1)$ and $\bs{c}=(179,1,\dotsc,1)$ for a $240\times 301$ matrix.  By conditioning on the entry in the first row and the first column and then using symmetry, one can see that 
\[ N(\bs{r},\bs{c}) = \binom{300}{240}\binom{239}{179}60! + \binom{300}{239}\binom{239}{178}61! \approx 9.6843103\times 10^{205} \]
Generating a single observation takes about 0.077 s. 
\eqref{e:g2} and \eqref{e:g2old} have $W_k=N(\bs{r},\bs{c})$ for all $k$ since they are exactly uniform, giving $\widehat\Delta = 1$, $\widehat{cv}{}^2=0$ (where we take $0/0=0$) and $\overline W \pm S_{\overline W}=N(\bs{r},\bs{c})\pm 0$.  (The equivalence is not truly exact because of rounding errors; in this case $\overline W$ agreed with $N(\bs{r},\bs{c})$ to 12 digits.)  On the other hand, using $N=10^5$ samples, \eqref{e:g1} gives $\widehat\Delta = 4.1\times 10^{11}$, $\widehat{cv}{}^2=1.7\times 10^3$, and $\overline W \pm S_{\overline W} = (2.2\pm 0.3)\times 10^{205}$.  Similarly, \eqref{e:g1old} gives $\widehat\Delta = 5.9\times 10^{17}$, $\widehat{cv}{}^2=4.5\times 10^3$, and $\overline W \pm S_{\overline W} = (4.4\pm 0.9)\times 10^{203}$. Both are clearly far from uniform and approximate 95\% confidence intervals of the form $\overline W \pm 2S_{\overline W}$ do not contain $N(\bs{r},\bs{c})$.  For \eqref{e:g1old}, \cite{bezakova2006nes} point out that ignoring the importance sampling diagnostics and relying only on the apparent convergence of $\overline W$ can be quite misleading.  They also show that true convergence of $\overline W$ requires $N$ to be exponential in the size of the margins.

Finally, consider Darwin's finch data (see CDHL) which is a $13\times 17$ occurrence matrix with $\bs{r}:=$ (14, 13, 14, 10, 12,  2, 10, 1, 10, 11,  6,  2, 17) and $\bs{c}:=$ (4,  4, 11, 10,  10,  8,   9,   10, 8, 9,    3,  10,  4,   7,   9,    3,   3).  A single sample takes about $0.001$ s.  Table \ref{t:finch} summarizes the performance based on $N=10^6$ samples.  For comparison, CDHL report that $N(\bs{r},\bs{c}) = $ 67,149,106,137,567,626 for the finch data.  This example is interesting because none of the algorithms perform well, even though it is quite small, emphasizing the fact that the heuristics for choosing $\bs{p}$ are motivated by asymptotic approximations and may perform poorly on small problems.

\begin{table}[h]
\caption{Performance for $N=10^6$ on Darwin's finch data.}
\begin{tabular}{|c||c|c|c|} \hline
$\vphantom{\hat{\bigl(}}$ & $\widehat\Delta$ & $\widehat{cv}{}^2$ & $\overline W \pm S_{\overline W}$  \\ [1ex] \hline \hline
\eqref{e:g1} $\vphantom{\bigl(}$ & $2.8\times 10^3$ & 0.4363 & $(6.722 \pm 0.004) \times 10^{16}$ \\ \hline
\eqref{e:g1old} $\vphantom{\bigl(}$ & $5.6\times 10^4$ & 1.1467 & $(6.715 \pm 0.007) \times 10^{16}$ \\ \hline
\eqref{e:g2} $\vphantom{\bigl(}$ & $1.0\times 10^5$ & 1.3710 & $(6.727 \pm 0.008) \times 10^{16}$ \\ \hline
\eqref{e:g2old} $\vphantom{\bigl(}$ & $5.4\times 10^6$ & 4.0081 & $(6.729 \pm 0.013) \times 10^{16}$ \\ \hline
\end{tabular}
\label{t:finch}
\end{table}

These experiments suggest that \eqref{e:g1} is a good default choice, except for very sparse and/or irregular margins, where \eqref{e:g2} works better.  (In the neuroscience application that motivated this research, the margins are sparse and irregular and \eqref{e:g2} has far superior performance over \eqref{e:g1}.)

\subsection{External checks on uniformity}

The previous experiments are based primarily on the internal diagnostics of samples from the proposal distribution $Q$.  Other than \cite{blanchet2006isa}'s analysis of \eqref{e:g2old} (which is not even one of the algorithms we finally recommend) and the fact that $\bs{p}$ is based on sensible heuristics, there are no external checks on the uniformity of $Q$.  Using a complicated, high dimensional proposal distribution without external checks can be dangerous.  Indeed, consider the following worst-case scenario.  Suppose that $\Omega(\bs{r},\bs{c})=E\cup E^c$, where $E$ is much smaller than $E^c$, and suppose that $Q$ is uniform over each of $E$ and $E^c$, but far from uniform over $\Omega(\bs{r},\bs{c})$, namely,
\[ Q(\bs{z}) = \frac{1-\epsilon}{|E|}\ind\{\bs{z}\in E\} + \frac{\epsilon}{|E^c|}\ind\{\bs{z}\in E^c\} \quad  \text{for} \ \frac{|E|}{|E^c|} \ll \epsilon \ll 1 \]
If $\epsilon$ is extremely tiny, say $\epsilon = 10^{-100}$, then Monte Carlo samples from $Q$ will (practically speaking) always lie in $E$, which itself is a tiny fraction of $\Omega(\bs{r},\bs{c})$.  Furthermore, all internal diagnostics will report that $Q$ is exactly uniform, since it is uniform over $E$.  But, of course, statistical inferences based on samples from $Q$ will tend to be completely wrong.  This section describes two types of experiments designed to provide external checks on the uniformity of $Q$. 

For the first set of experiments we generate a binary matrix $\bs{Z}$ from the uniform distribution over all binary matrices with row sums $\bs{r}$.  This is easy to do by independently and uniformly choosing each row of $\bs{Z}$ from one of the $\binom{n}{r_i}$ possible configurations.  Since the conditional distribution of $\bs{Z}$ given its columns sums $\bs{C}$ is uniform over $\Omega(\bs{r},\bs{C})$, we can view $\bs{Z}$ as a single observation from the uniform distribution over $\Omega(\bs{r},\bs{C})$.  (Of course, there is no practical way to uniformly and independently generate another such $\bs{Z}$ with the same $\bs{C}$.)  Notice that $Q(\bs{Z})$ gives external information about the uniformity of $Q$ for these margins, since it gives the value of $Q$ at a uniformly chosen location in $\Omega(\bs{r},\bs{C})$.  Indeed, in the pathological thought experiment described above, $\bs{Z}$ would almost certainly be in $E^c$ and $Q(\bs{Z})^{-1}$ would be substantially larger than any of the importance weights.  Alternatively, if $Q$ is nearly uniform, then $Q(\bs{Z})^{-1}$ should be indistinguishable from the other importance weights.  In summary, we can compare $Q$ to $P$ by comparing the importance weights to $Q(\bs{Z})^{-1}$.  (This observation can also be used to give valid Monte Carlo p-values with importance sampling, even if the importance sampling distribution is far from the target distribution.) 

Each experiment of this type proceeds identically.  We fix $m$, $n$, and $\bs{r}$.  Then we generate $L$ iid observations, say $\bs{Z_0^1},\dotsc,\bs{Z_0^L}$, from the uniform distribution over all $m\times n$ binary matrices with row sums $\bs{r}$.  The column sums of these matrices are $\bs{C^1},\dotsc,\bs{C^L}$.  Then, for each $\ell=1,\dotsc,L$, we generate $N$ iid observations, say $\bs{Z^\ell_1},\dotsc,\bs{Z^\ell_N}$, from the proposal distribution $Q$ over $\Omega(\bs{r},\bs{C^\ell})$.  We compute the ratio of maximum to minimum importance weights {\em including the original observation} for each $\ell$, namely, 
\[ \widehat\Delta^\ell := \frac{\max_{k=0,\dotsc,N} Q(\bs{Z^\ell_k})^{-1}}{\min_{k=0,\dotsc,N} Q(\bs{Z^\ell_k})^{-1}} \]
and we report the final summary $\widehat\Delta_{\max}  := \max_{\ell=1,\dots,L} \widehat\Delta^\ell$.  We do this for each of the four heuristics for choosing $\bs{p}$ (using the same $\bs{Z_0^1},\dotsc,\bs{Z_0^L}$ for each).  If $\widehat\Delta_{\max}$ is close to one, then this provides evidence that $Q$ is approximately uniform over a large part of each of the $\Omega(\bs{r},\bs{C^\ell})$'s. 

We begin with regular row sums as in table \ref{t:WN1_1}.  Take $m=n=1000$ and $r_1=r_i$ for all $i$.  We use $L=10$ and $N=10$ for the cases $r_1=2,8,32$.  Table \ref{t:WN1_1*} reports $\widehat\Delta_{\max}$.  The results are very close to uniform. 

\begin{table}[h]
\caption{${\widehat\Delta}_{\max}$ for $m=n=1000$ and $r_1=r_i$ for all $i$.}
\label{t:WN1_1*}
\begin{tabular}{|c||c|c|c|} \hline
$r_1$ & 2 & 8 & 32 \\ \hline \hline
\eqref{e:g1} & 1.0002 & 1.0023 & 1.0051 \\ \hline
\eqref{e:g1old} & 1.0880 & 1.2130 & 1.3145 \\ \hline
\eqref{e:g2} & 1.0000 & 1.0011 & 1.0845 \\ \hline
\eqref{e:g2old} & 1.0001 & 1.0086 & 1.7087 \\ \hline
\end{tabular}
\end{table}

For another example, take the row sums for the irregular $50\times 100$ case that was used for table \ref{t:WN1_2} and take $k=1$, i.e., $\bs{r}=\bs{\tilde r}$.   We use $L=100$ and $N=1000$ and find that $\widehat\Delta_{\max}=$ 2.264,   25.42,    1.875, and    7.962 for \eqref{e:g1}, \eqref{e:g1old}, \eqref{e:g2}, and \eqref{e:g2old}, respectively.    

Taken with the results in the previous section these preliminary experiments suggest that \eqref{e:g1} and \eqref{e:g2} are somewhat uniform (certainly they are not pathological) over a large part of the sample space for margins like the ones considered here.  Being somewhat uniform over most of the sample space is the primary requirement for a useful proposal distribution if the ideal goal is to sample uniformly over the sample space.  

For the second type of experiment, we try to design an extreme $\bs{z^*}\in\Omega(\bs{r},\bs{c})$ and compare the importance weights to $Q(\bs{z^*})^{-1}$.  Again, if $Q$ is approximately uniform over all of $\Omega(\bs{r},\bs{c})$ then $Q(\bs{z^*})^{-1}$ should be indistinguishable from the other importance weights.  For these experiments we report
\[ {\widehat\Delta}^* := \frac{\max\left\{Q(\bs{z^*})^{-1},W_1,\dotsc,W_N\right\}}{\min\left\{Q(\bs{z^*})^{-1},W_1,\dotsc,W_N\right\}} \]
which should be close to one if the region in $\Omega(\bs{r},\bs{c})$ where $Q$ is approximately uniform includes $\bs{z^*}$.

Consider the regular case where $m=n=1000$ and $r_1=r_i=c_j$ for all $i,j$.  Suppose that $r_1$ evenly divides 1000 and let $\bs{z^*}$ be comprised only of disjoint $r_1\times r_1$ blocks of ones.  In particular, take $z^*_{ij}=1$ for $(k-1)r_1+1 \leq i,j \leq kr_1$ and for $k=1,\dotsc,1000/r_1$.  For each of the four $\bs{p}$ heuristics and for the cases $r_1=2,4,8$ we compute $Q(\bs{z^*})^{-1}$ and compare it to the data that generated the corresponding parts of table \ref{t:WN1_1}.  Table \ref{t:z*_1} summarizes the results with ${\widehat\Delta}^*$ defined above.
The departure from uniformity is striking, especially for \eqref{e:g1} which performed spectacularly based on internal diagnostics.  \eqref{e:g2} seems to be the best by this measure.  Clearly, these algorithms are not within a few percent of uniform over {\em all} of $\Omega(\bs{r},\bs{c})$ as tables \ref{t:WN1_1} and \ref{t:WN1_1*} might seem to suggest.

\begin{table}[h]
\caption{${\widehat\Delta}^*$ for the corresponding part of Table \ref{t:WN1_1}.}
\label{t:z*_1}
\begin{tabular}{|c||c|c|c|} \hline
$r_1$ & 2 & 4 & 8  \\ \hline \hline
\eqref{e:g1} $\vphantom{\bigl(}$  & 1.741   &   27.24  &  $6.25\times 10^4$ \\ \hline
\eqref{e:g1old} $\vphantom{\bigl(}$  & 43.03   &    $1.14\times 10^5$  & $1.24\times 10^{12}$ \\ \hline
\eqref{e:g2} $\vphantom{\bigl(}$  & 1.367   &   1.567   &   197.3 \\ \hline
\eqref{e:g2old} $\vphantom{\bigl(}$  & 1.353   &    4.115 &  $4.52\times 10^7$ \\ \hline
\end{tabular}
\end{table}

For another example, consider the irregular $50\times 100$ margins that were used for table \ref{t:WN1_2} and take $k=1$, i.e., $\bs{r}=\bs{\tilde r}$ and $\bs{c}=\bs{\tilde c}$.  We construct a pathological $\bs{z^*}$ as follows.  Place $c_1$ ones in the last $c_1$ rows (corresponding to the smallest row sums) of the first column.  Place $c_2$ ones in the last available $c_2$ rows of the second column, where a row is available if placing a one in that row will not exceed the row sum for that row.  Continue in this manner until all the columns are assigned or until a column cannot be assigned successfully.  In general, this procedure will not terminate successfully, but it does for this choice of margins.  The resulting $\bs{z^*}$ is unusual because rows and columns with large sums tend to have zeros at the intersecting entry.  Using the data from the corresponding part of Table \ref{t:WN1_2} gives ${\widehat\Delta}^* = 15.37$, $2.1\times 10^5$, $4.7\times 10^3$, and $4.8\times 10^8$ for \eqref{e:g1}, \eqref{e:g1old}, \eqref{e:g2}, and \eqref{e:g2old}, respectively.  Again, this shows that $Q$ is not uniformly uniform over $\Omega(\bs{r},\bs{c})$.  The departure from uniformity is substantial enough that $Q$ can only be used for importance sampling, and not, for example, as a proposal distribution for (efficient) rejection sampling to sample exactly from $P$.

\section{Exactly enforcing a zero diagonal}

In general, if we want to force certain entries to be either zero or one, then we can simply set the corresponding $A_i$ from Theorem \ref{t} to be either $\{0\}$ or $\{1\}$, respectively, in the dynamic programming algorithm above.  This will generate matrices with the correct margins and with the desired forced entries.  However, this may lead to situations where it is impossible to choose a valid column during the recursive generation of columns, since the margin constraints (in $\bs{B}$) do not account for the forced entries.  If the probability of failure is high, then $Q$ will not be a useful proposal distribution.  An interesting question is whether the margin constraints can be modified to account for forced entries so that $Q$ always generates valid matrices.  \cite{chen2007cit} investigates this question for the CDHL procedure and we do the same here, albeit for a very small class of forced entries --- essentially, the case of forcing a zero diagonal. 

Square binary matrices are often used to represent the adjacency matrix of a directed graph.  When these graphs cannot have self connections, the diagonal entries should be zero.  \cite{chen2007cit} gives a version of Theorem \ref{t} for this case and modifies the CDHL procedure accordingly.  \cite{greenhill2007aed} and \cite{Bender:Asymptotic:1974} provide the necessary asymptotics.  The results are slightly more general than square matrices with zero diagonal --- the only constraint is that each row and column can have at most one forced zero entry.  

Let $\bs a$ be a fixed and known $m\times n$ binary matrix and define
\[ \Omega(\bs{r},\bs{c},\bs{a}) := \bigl\{\bs{z}\in\Omega(\bs{r},\bs{c}) : z_{ij}a_{ij}=0, \ \forall i,j \bigr\} \]
to be those $m\times n$ binary matrices with margins $\bs{r}$ and $\bs{c}$ and with zeros at any location that $\bs{a}$ has a one.  These forced zeros are called {\em structural zeros} and the entries of $\bs{a}$ indicate their locations.  $N(\bs{r},\bs{c},\bs{a})$ and $\Omega_1(\bs{r},\bs{c},\bs{a})$ are defined similarly as before.  $\widetilde N(\bs{r},\bs{c},\bs{a})$ denotes a generic approximation to $N(\bs{r},\bs{c},\bs{a})$.  Specific suggestions can be found in Section \ref{s:Na} below. 

Following exactly our earlier development, the target distribution, $P$, is the uniform distribution over $\Omega(\bs{r},\bs{c},\bs{a})$ and the proposal distribution, $Q$, will be defined recursively over columns via
\[ Q_1(\bs{b}) \propto \widetilde N(\bs{r}-\bs{b},\bs{c'},\bs{a'})\ind\bigl\{\bs{b}\in\Omega_1(\bs{r},\bs{c},\bs{a})\bigl\}  \]
where $\bs{a'}$ is the $m\times(n-1)$ matrix created by removing the first column from $\bs{a}$, namely,
\[ a'_{ij} := a_{i,j+1} \quad \text{for } 1\leq i \leq m, \ 1 \leq j \leq n-1 . \]  
If $\bs{a}$ has at most one entry of one in each row and column (indicating at most one zero in each row and column), then the factorization in \eqref{e:Qf} continues to hold (at least for the suggestions for $\widetilde N$ described below), where the constants $p_1,\dotsc,p_m$ and the sets $A_1,\dotsc,A_m$ and $B_1,\dotsc,B_m$ depend on $\bs{r}$, $\bs{c}$, and $\bs{a}$ and are easy to compute.  The algorithm described in Section \ref{s:overview} can be used exactly, with the only differences being modified choices of $\bs{p}$, $\bs{A}$, and $\bs{B}$ in order to account for the forced zeros indicated by $\bs{a}$, and also a few additional constraints on the order in which the rows and columns are sampled.

In the following sections we define $\bs{\xi}$ and $\bs{\zeta}$ to be the row and column sums of $\bs{a}$, respectively, so that 
\[ \begin{aligned} \xi_i & = \text{\# of structural zeros in the $i$th row,} \\ \zeta_j & = \text{\# of structural zeros in the $j$th column}. \end{aligned} \]
Similarly, we define $\bs{\xi'}$ and $\bs{\zeta'}$ to be the margins of $\bs{a'}$.
As mentioned before, the final sampling algorithm relies on the assumption that 
\begin{equation} \label{e:1a} \xi_i \leq 1, \ \zeta_j\leq 1, \ \forall i,j \end{equation}
Although quite restrictive, this includes the important case of enforcing a zero diagonal.

\subsection{$N(r,c,a)$} \label{s:Na}

We first present the asymptotic enumeration results corresponding to Section \ref{s:N}.  Each of the suggestions \eqref{e:g1}, \eqref{e:g1old}, and \eqref{e:g2old} for computing $\bs{p}$ has a corresponding correction that accounts for structural zeros.  These corrections allow for arbitrary structural zeros and do not rely on the constraint that each row and column has at most one structural zero.  The computational cost for the corrections is negligible.  Lemma \ref{l:f} becomes
\begin{lem} \label{l:fa}
Suppose that
\begin{equation} \widetilde N(\bs{r},\bs{c},\bs{a}) =  g_0(\bs{c},\bs{a},m,n)\prod_{i=1}^{m} g_i(r_i,\bs{c},\bs{a},m,n) \label{e:lemfa} \end{equation}
whenever $N(\bs{r},\bs{c},\bs{a}) > 0$ for some functions $(g_i:i=0,\dotsc,m)$, where $m$ and $n$ are the respective number of elements of $\bs{r}$ and $\bs{c}$.  Then as $\bs{b}$ varies over $\Omega_1(\bs{r},\bs{c},\bs{a})$, we have
\begin{equation} \label{e:Nfa} \widetilde N(\bs{r}-\bs{b},\bs{c'},\bs{a'}) \propto \prod_{i=1}^m p_i^{b_i}(1-p_i)^{1-b_i} \end{equation}
where 
\begin{equation} \label{e:pia} p_i := \frac{g_i(r_i-1,\bs{c'},\bs{a'},m,n-1)}{g_i(r_i,\bs{c'},\bs{a'},m,n-1)+g_i(r_i-1,\bs{c'},\bs{a'},m,n-1)} . \end{equation}
\end{lem}

\cite{greenhill2007aed} provide the modified asymptotic enumeration results corresponding to \eqref{e:n1} and \eqref{e:g1}.
\begin{gather} \label{e:n1a}
\begin{split}
\widetilde N(\bs{r},\bs{c},\bs{a}) := \binom{mn-[\bs{\zeta}]_1}{[\bs{c}]_1}^{-1}\prod_{i=1}^m \binom{n-\xi_i}{r_i}\prod_{j=1}^n \binom{m-\zeta_j}{c_j} \\ \times \exp\left(-\frac{1}{2}\bigl(1-\mu\bigr)\bigl(1-\nu\bigr)-\eta\right) \end{split} \\
\mu := \mu(\bs{r},\bs{c},m,n) := \frac{mn}{[\bs{c}]_1\bigl(mn-[\bs{c}]_1\bigr)}\sum_{i=1}^m \bigl(r_i-[\bs{c}]_1/m\bigr)^2  \notag \\
\nu := \nu(\bs{c},m,n) := \frac{mn}{[\bs{c}]_1\bigl(mn-[\bs{c}]_1\bigr)}\sum_{j=1}^n \bigl(c_j-[\bs{c}]_1/n\bigr)^2 \notag \\
\eta := \eta(\bs{r},\bs{c},\bs{a},m,n) := \frac{mn}{[\bs{c}]_1\bigl(mn-[\bs{c}]_1\bigr)}\sum_{i=1}^m\sum_{j=1}^n a_{ij}\bigl(r_i-[\bs{c}]_1/m\bigr)\bigl(c_j-[\bs{c}]_1/n\bigr) \notag
\end{gather}
which permits the factorization in Lemma \ref{l:fa} with
\[ \begin{split} g_i(r_i,\bs{c},\bs{a},m,n) := \binom{n-\xi_i}{r_i}\exp\biggl(\frac{mn\bigl(r_i-[\bs{c}]_1/m\bigr)^2}{2[\bs{c}]_1\bigl(mn-[\bs{c}]_1\bigr)}\bigl(1-\nu(\bs{c},m,n)\bigr) \\ -\frac{mnr_i}{[\bs{c}]_1\bigl(mn-[\bs{c}]_1\bigr)}\sum_{j=1}^na_{ij}\bigl(c_j-[\bs{c}]_1/n\bigr)\biggr) \end{split} \] 
giving the Bernoulli probabilities
\begin{subequations}
\begin{gather} p_i := \frac{r_i\exp\Bigl(\beta\bigl(1-2(r_i-[\bs{c'}]_1/m)\bigr)+\delta_i\Bigr)}{n-\xi_i'-r_i+r_i\exp\Bigl(\beta\bigl(1-2(r_i-[\bs{c'}]_1/m)\bigr)+\delta_i\Bigr)}  \label{e:g1a} 
\\ \notag \beta := \frac{m(n-1)}{2[\bs{c'}]_1\bigl(m(n-1)-[\bs{c'}]_1\bigr)}\bigl(1-\nu(\bs{c'},m,n-1)\bigr)  
\\ \notag \delta_i := \frac{m(n-1)}{[\bs{c'}]_1\bigl(m(n-1)-[\bs{c'}]_1\bigr)}\sum_{j=1}^{n-1}a'_{ij}\bigr(c'_j-[\bs{c'}]_1/(n-1)\bigl)
\end{gather}
where we take $0/0:=0$.
This is a refinement over the suggestion in \cite{chen2007cit} to use $g_i(r_i,\bs{c},\bs{a},m,n) := \binom{n-\xi_i}{r_i}$ and leading to 
\begin{equation}  p_i := \frac{r_i}{n-\xi'_i}  . \label{e:g1olda} \end{equation}
\end{subequations}

Modifications of \eqref{e:n2} and the corresponding $\bs{p}$ in \eqref{e:g2} are not currently available.  For the approximation that led to \eqref{e:g2old}, however, \cite{Bender:Asymptotic:1974} and \cite{mckay1984amp} give the necessary modification for structural zeros: 
\begin{equation} \label{e:n2olda}
\widetilde N(\bs{r},\bs{c},\bs{a}) := \frac{[\bs{c}]_1!}{\prod_{i=1}^m r_i!\prod_{j=1}^nc_j!}\exp\left(-\frac{[\bs{r}]_2[\bs{c}]_2}{2[\bs{c}]_1^2}-\frac{\sum_{i=1}^m\sum_{j=1}^n a_{ij} r_ic_j}{[\bs{c}]_1}\right)
\end{equation}
which permits the factorization in Lemma \ref{l:fa} with
\[  g_i(r_i,\bs{c},\bs{a},m,m) := \frac{1}{r_i!}\exp\left(-\frac{[r_i]_2[\bs{c}]_2}{2[\bs{c}]_1^2}-\frac{r_i\sum_{j=1}^na_{ij}c_j}{[\bs{c}]_1}
\right) 
\]
\eqref{e:pia} gives the Bernoulli probabilities
\begin{equation} \label{e:g2olda}
p_i := \frac{r_i\exp\bigl((r_i-1)[\bs{c'}]_2/[\bs{c'}]_1^2+\sum_{j=1}^{n-1}a'_{ij}c'_j/[\bs{c}]_1\bigr)}{1+r_i\exp\bigl((r_i-1)[\bs{c'}]_2/[\bs{c'}]_1^2+\sum_{j=1}^{n-1}a'_{ij}c'_j/[\bs{c}]_1\bigr)}
\end{equation}
where we take $0/0:=0$.
\cite{chen2007cit} also noted this approximation, but did not use it in any examples.

Finally, we remark again that any of the approximations in Section \ref{s:N} could also be used in the case with structural zeros.  Ignoring the zeros in the choice of $\bs{p}$ will presumably cause a drop in performance, but this may be negligible, especially for large matrices.  In particular, \eqref{e:g2} may still be quite useful for large sparse irregular matrices with structural zeros.

\subsection{$\Omega_1(r,c,a)$} 
  
Here we describe the discrete mathematics corresponding to Section \ref{s:O}.  The goal is to show that $\Omega_1(\bs{r},\bs{c},\bs{a})$ factors like \eqref{e:Omega}.  This factorization relies on the assumption that $\bs{a}$ encodes at most one forced zero in each row and column.  
\begin{thm} \label{t:a} \cite{chen2007cit} Fix an $m\times n$ binary matrix $\bs{a}$ with row sums $\bs{\xi}$ and column sums $\bs{\zeta}$ such that $\max_i\xi_i\leq 1$ and $\max_j\zeta_j\leq 1$.  Define the location of the structural zero in each row by
\[ y_i = \begin{cases} \text{the unique $j$ such that $a_{ij}=1$} & \text{if $\xi_i = 1$} \\ n+1 & \text{if $\xi_i = 0$} \end{cases} \]
Assume $N(\bs{r},\bs{c},\bs{a}) > 0$ and assume that the columns and rows are ordered so that $c_1 \geq \dotsb \geq c_n$ and $r_1\geq \dotsb \geq r_m$ with the further constraint that $r_i=r_{i+1}$ implies that $y_i\leq y_{i+1}$.
Define $\bs{A}:=A_1\times\dotsb\times A_m$ and $\bs{B}:=B_1\times\dotsb\times B_m$ by
\begin{align*} A_i & := \begin{cases} \{0\} & \text{if $r_i=0$ or $a_{i1}=1$;} \\ \{0,1\} & \text{if $0 < r_i < \textstyle n-\xi_i$ and $a_{i1}=0$;} \\
\{1\} & \text{if $r_i=\textstyle n-\xi_i$ and $a_{i1}=0$;} \end{cases} \\
B_i & := \begin{cases} \textstyle \Bigl\{\max\bigl\{0,\bigl(\sum_{\ell=1}^i r_\ell\bigr)-d_i\bigr\} , \dotsc, c_1\Bigr\} & \text{if $i < m$;} \\
\{c_1\} & \text{if $i = m$;} \end{cases} 
\end{align*}
where
\[ d_i := \min_{j=1,\dotsc,n}    \bigl[\textstyle i(j-1)   +\sum_{k=j+1}^n c_k   -\sum_{\ell=1}^i\sum_{k=2}^j a_{\ell k}\bigr]  \]
Let $\bs{b}$ be a binary $m$-vector and let $\bs{s}$ denote the partial sums of $\bs{b}$ defined by $s_i:=\sum_{\ell=1}^i b_i$.   Then $\bs{b}\in\Omega_1(\bs{r},\bs{c},\bs{a})$ if and only if $\bs{b}\in \bs{A}$ and $\bs{s}\in \bs{B}$.  In other words,
\begin{equation}
\ind\bigl\{\bs{b}\in\Omega_1(\bs{r},\bs{c},\bs{a})\bigr\} = \prod_{i=1}^m \ind\{b_i\in A_i\}\ind\bigl\{\textstyle\sum_{\ell=1}^i b_\ell \in B_i\bigr\}
\end{equation} 
\end{thm}

Sampling proceeds exactly as before.  Simply replace Theorem \ref{t} with Theorem \ref{t:a} to compute the sets $\bs{A}$ and $\bs{B}$.  The only differences are that the columns must be sampled in order of decreasing column sums --- we did this for efficiency in the examples above, but now it is imperative --- and the structural zeros can affect how ties are decided when ordering the rows.   The additional computation required for precomputing $d_k$ is negligible.  

\section{Conclusions}

This paper improves upon the algorithms in \cite{Chen:Sequential:2005} for approximate uniform generation of binary matrices with specified margins.  Such algorithms are useful  for Monte Carlo approximate inference and counting (via importance sampling).  An important aspect of the improvement here is the use of dynamic programming (DP) to exactly enforce the margin constraints.  The appeal of the DP perspective is that it immediately suggests a variety of generalizations, such as, arbitrary structural zeros, symmetric matrices and integer-valued matrices.
The primary challenges are to
\begin{itemize}
\item[(a)] formulate the constraints into a form that DP can use;
\item[(b)] find approximate counting formulas that are easy to compute;
\end{itemize} 
If part (a) fails, it may be possible to formulate a subset of the constraints into a form that DP can use.  As long as the support of the resulting $Q$ is not too much larger than the support of $P$, importance sampling can still work well.  Part (b) remains an important challenge.  Fortunately, in the case considered here the approximations in \cite{greenhill2006aes} and \cite{canfield2008aed} are easy to compute and work well on most examples.  As improved approximations for $N(\bs{r},\bs{c})$ appear in the literature they can be used immediately via \eqref{e:pi2} to create improved proposal distributions.

For the two approximations used here, namely the ones leading to \eqref{e:g1} and \eqref{e:g2}, the resulting proposal distributions seem to be approximately uniform over a large part of the sample space for a wide class of margins, with \eqref{e:g1} working well when the margins are close to regular and \eqref{e:g2} working well when the margins are sparse.  The case of highly irregular and dense margins does not seem to be handled well by either algorithm.  The algorithms are not, however, uniformly uniform --- there are regions (albeit very small regions) with substantially smaller proposal probabilities --- so these proposal distributions cannot be used for efficient exact uniform sampling via rejection sampling.  The ideal goal of fast exact uniform sampling from $\Omega(\bs{r},\bs{c})$ remains elusive.

\appendix

\section{Importance sampling} \label{s:is}

This section contains a description of importance sampling as it relates to the problem at hand.  We want to generate a random object $\bs{Z}$ from a {\em target distribution} $P$.    In the main text $\bs{Z}$ is an $m\times n$ binary matrix and $P$ is the uniform distribution conditioned on certain row and column sums.   Usually, the goal is to generate an independent and identically distributed (i.i.d.)~sequence $\bs{Z_1},\bs{Z_2},\dotsc,\bs{Z_N}$ with common distribution $P$ for the purposes of Monte Carlo inference.  In particular, we can Monte Carlo approximate the expected value under $P$ (denoted $E_P$) of any function $f$ via the strong law of large numbers, namely, 
\begin{equation} \label{e:mcnaive} \frac{1}{N}\sum_{k=1}^n f(\bs{Z_k}) \to E_P[f(\bs{Z})] \end{equation}
almost surely as $N\to\infty$.    

For many purposes, though, it is enough to be able to sample from a different distribution $Q$, called the {\em proposal distribution}, for which the probability of samples are easy to evaluate and whose support contains the support of $P$, the target distribution.  Expectations under $P$ are related to expectations under $Q$ via
\begin{align*} E_P[f(\bs{Z})] & = \sum_{\bs z} f(\bs{z})P(\bs{z}) = \sum_{\bs z} f(\bs{z})P(\bs{z})Q(\bs{z})^{-1}Q(\bs{z}) \\ & = E_Q\bigl[f(\bs{Z})P(\bs{Z})Q(\bs{Z})^{-1}\bigr] \end{align*}
where $E_P$ is the expected value when $\bs{Z}$ has the target distribution $P$ and $E_Q$ is the expected value when $\bs{Z}$ has the proposal distribution $Q$.  This implies that we can generate $\bs{Z_1},\bs{Z_2},\dotsc,\bs{Z_N}$ i.i.d.~from the {\em proposal} distribution $Q$ and Monte Carlo approximate $E_P[f(\bs{Z})]$ using
\begin{equation} \frac{1}{N}\sum_{k=1}^n f(\bs{Z_k})P(\bs{Z_k})Q^{-1}(\bs{Z_k}) \to E_Q\bigl[f(\bs{Z})P(\bs{Z})Q(\bs{Z})^{-1}\bigr] = E_P[f(\bs{Z})] \label{e:mcis} \end{equation}
where the convergence happens almost surely as $N\to\infty$.  This is called {\em importance sampling} and usually the goal is to choose $Q$ so that $f(\bs{Z})P(\bs{Z})Q^{-1}(\bs{Z})$ has small variance when $\bs{Z}$ has distribution $Q$.  If this variance is smaller than the variance of $f(\bs{Z})$ when $\bs Z$ has distribution $P$, then importance sampling will need fewer Monte Carlo samples than the original exact sampling because the convergence in \eqref{e:mcis} will happen faster than the convergence in \eqref{e:mcnaive}.

In our situation, however, it turns out that $P$ is known only up to a constant of proportionality, say
\[ P(\bs{z}) = \kappa^{-1} R(\bs{z}) \]
where $R$ is known, but $\kappa$ is not.  (In our case, $P$ will be uniform and so $R$ is simply the indicator function of the support of $P$.)  Importance sampling can still be used, but now $\kappa$ must also be related to expectations under $Q$ via
\[ \kappa = \sum_{\bs z} R(\bs{z}) = \sum_{\bs z} R(\bs{z})Q(\bs{z})^{-1}Q(\bs{z}) = E_Q\bigl[R(\bs{Z})Q(\bs{Z})^{-1}\bigr] \]  
and estimated with
\begin{equation} \frac{1}{N}\sum_{k=1}^n R(\bs{Z_k})Q^{-1}(\bs{Z_k}) \to E_Q\bigl[R(\bs{Z})Q(\bs{Z})^{-1}\bigr] = \kappa \label{e:mciskappa} \end{equation}
Notice that efficient estimation of $\kappa$ requires that $R(\bs{Z})Q(\bs{Z})^{-1}$ have small variance when $\bs{Z}$ has distribution $Q$.  In other words, {\em we need the proposal distribution to be very similar to the target distribution}.  This is quite different from the typical goal of importance sampling.

We can now relate expectations under $P$ and $Q$ via
\begin{align} E_P[f(\bs{Z})] & = E_Q\bigl[f(\bs{Z})P(\bs{Z})Q(\bs{Z})^{-1}\bigr] = \kappa^{-1}E_Q\bigl[f(\bs{Z})R(\bs{Z})Q(\bs{Z})^{-1}\bigr] \notag \\ & = \frac{E_Q\bigl[f(\bs{Z})R(\bs{Z})Q(\bs{Z})^{-1}\bigr]}{E_Q\bigl[R(\bs{Z})Q(\bs{Z})^{-1}\bigr]} \notag \end{align}
and consistent Monte Carlo estimation proceeds via
\begin{equation} \frac{N^{-1}\sum_{k=1}^n f(\bs{Z_k})R(\bs{Z_k})Q^{-1}(\bs{Z_k})}{N^{-1}\sum_{k=1}^n R(\bs{Z_k})Q^{-1}(\bs{Z_k})} \to \frac{E_Q\bigl[f(\bs{Z})R(\bs{Z})Q(\bs{Z})^{-1}\bigr]}{E_Q\bigl[R(\bs{Z})Q(\bs{Z})^{-1}\bigr]} = E_P[f(\bs{Z})] \label{e:mcis*} \end{equation}
where, again, $\bs{Z_1},\bs{Z_2},\dotsc,\bs{Z_N}$ are i.i.d.~$Q$.
This procedure will be most efficient when the variability of both the numerator and the denominator are small.  These may be competing demands.  We do not address the challenging problem of designing a proposal distribution that balances variability in both the numerator and denominator.  Rather, in the spirit of CDHL we merely try to reduce variability in the denominator by designing a proposal distribution that is as close to the target as possible.

Choosing the proposal to closely match the target (and ignoring the function $f$) may seem like a strange goal when viewed from the typical importance sampling perspective.  Indeed, perfectly achieving this goal makes the importance sampling approach in \eqref{e:mcis*} reduce to the exact sampling approach in \eqref{e:mcnaive}.  So we might expect that for many functions $f$, importance sampling is less efficient than exact sampling.  However, suppose that we can generate observations from $Q$ much faster than observations from $P$.  Then the total computational time for importance sampling might still be faster than for exact sampling, even though larger sample sizes are required for importance sampling.  This observation was the motivation for CDHL and it is the motivation here: Exact uniform sampling of binary matrices with margin constraints is computationally prohibitive; but approximate uniform sampling can be quite fast.

It is also important to keep in mind that the exact sampling Monte Carlo approximation in \eqref{e:mcnaive} would be fine in many situations {\em if we could sample efficiently from $P$}.  For example, in hypothesis testing $f$ is usually the indicator function of some rejection region.  While this may be a rare event, it is usually enough to determine that the probability is indeed small (say, smaller than $0.05$, for example), and it is not necessary to accurately estimate the log-probability (which might require importance sampling in the usual sense).  So having a fast procedure that closely matches the statistical efficiency of \eqref{e:mcnaive} is very desirable.

To summarize, our goal in the text is to design approximately uniform distributions over binary matrices with margin constraints that permit fast sampling algorithms.  Monte Carlo inference (if that is what we want to do) can proceed via the left-hand side of \eqref{e:mcis*}, where the function $f$ is chosen by the practitioner, where the function $R(\bs{z})$ will simply be the indicator function that $\bs{z}$ is a binary matrix with the appropriate margins, and where $Q$ denotes the approximately uniform distribution that generated the i.i.d.~Monte Carlo samples $\bs{Z_1},\bs{Z_2},\dotsc,\bs{Z_N}$.

\bibliographystyle{apalike}
\bibliography{cpsampling}

\end{document}